\documentclass[12pt,draftcls,onecolumn]{IEEEtran}

\usepackage{epsf}
\usepackage{graphics}
\usepackage{times}
\usepackage{graphicx}
\usepackage{amsmath}
\usepackage{amssymb}
\usepackage{epsfig}
\usepackage{subfigure}
\usepackage{array}
\usepackage{float}

\begin{document}
\title{Clipping Noise Cancellation for OFDM and OFDMA Systems Using Compressed Sensing}

\author{Kee-Hoon~Kim,~Hosung~Park,~Jong-Seon~No,~and~Habong~Chung

\thanks{K.-H. Kim, H. Park, and J.-S. No are with the
Department of Electrical Engineering and Computer Science, INMC, Seoul
National University, Seoul, 151-744, Korea (phone: +82-2-880-8437,
fax: +82-2-880-8222, email: kkh@ccl.snu.ac.kr, lovepark98@ccl.snu.ac.kr, jsno@snu.ac.kr).}
\thanks{H. Chung is with the School of Electronics and Electrical Engineering,
Hong-Ik University, Seoul 121-791, Korea (e-mail: habchung@hongik.ac.kr).}

}

\markboth{IEEE TRANSACTIONS ON SIGNAL PROCESSING}%
{Shell \MakeLowercase{\textit{et al.}}: Bare Demo of IEEEtran.cls for Journals}

\maketitle

\begin{abstract}

In this paper, we propose clipping noise cancellation scheme using
compressed sensing (CS) for orthogonal frequency division
multiplexing (OFDM) systems. In the proposed scheme, only the data
tones with high reliability are exploited in reconstructing the
clipping noise instead of the whole data tones. For reconstructing
the clipping noise using a fraction of the data tones at the
receiver, the CS technique is applied. The proposed scheme is also
applicable to interleaved orthogonal frequency division multiple
access (OFDMA) systems due to the decomposition of fast Fourier
transform (FFT) structure. Numerical analysis shows that the
proposed scheme performs well for clipping noise cancellation of
both OFDM and OFDMA systems.
\end{abstract}


\section{Introduction}

\IEEEPARstart{O}{rthogonal} frequency division multiplexing (OFDM)
is known as one of the best modulation schemes for high rate data
transmission in wireless communication due to its robustness
against multipath fading, bandwidth efficiency, and simple
implementation. However, due to high peak-to-average power ratio
(PAPR) of the OFDM signals, the OFDM system requires expensive
high power amplifiers having a large dynamic range.

In the literatures, many PAPR reduction schemes for the OFDM
systems have been proposed \cite{Dae-Woon}-\cite{Ochiai}. The
simplest method for PAPR reduction of the OFDM signals is clipping
\cite{Xiaodong}. Clipping at the Nyquist sampling rate has been
used for low complexity applications but suffers from peak
regrowth after digital-to-analog (D/A) conversion. It is known
that clipping of the oversampled OFDM signals reduces the peak
regrowth after D/A conversion. But, it causes out-of-band
radiation which has to be filtered. Furthermore, clipping of OFDM
signals causes clipping noise which has sparsity in time domain.
There are several schemes to mitigate this clipping noise in each
of the two sampling rate cases \cite{Hangjun}-\cite{Ying}. The scheme in \cite{Hangjun} shows good clipping noise cancellation performance, but it requires iterative maximum likelihood (ML) estimation for all tones with clipping and filtering at the receiver, which causes lots of computation.

Compressed sensing (CS) is a sampling method that converts input
signal in high dimension into the signal lying in the smaller
dimension \cite{Donoho}-\cite{Wright}. In general, it is not
enough to recover an unknown signal using compressed observations
in the reduced dimension. Nevertheless, if the input signal has
sparsity, its reconstruction can be achieved at the receiver using
CS reconstruction algorithm. In this context, the clipping noise at the receiver can be
reconstructed by CS reconstruction algorithm. As the first work
for this, tone reservation to reconstruct the clipping noise by CS
is proposed in \cite{Al-Safadi}, where several
tones are reserved at the transmitter before clipping and the
receiver can reconstruct the clipping noise using CS reconstruction
algorithm. However in this scheme, the reserved tones induce data
rate loss and they should be demodulated at the receiver unlike
the conventional tone reservation scheme. Additionally, this
scheme shows poor bit error rate (BER) performance due to the
small number of compressed observations (i.e., the small number of reserved tones) and vulnerability of CS
reconstruction algorithm against noise. Motivated by \cite{Al-Safadi}, \cite{Mohammad} is proposed, where there is no data rate loss because pilot tones are exploited as compressed observations to reconstruct the clipping noise without reserved tones. But, it still shows poor BER performance for the same reasons as \cite{Al-Safadi}.

Orthogonal frequency division multiple access (OFDMA) has severer
PAPR problem due to the increased number of subcarriers.
There are several reasons that many clipping noise cancellation
schemes for OFDM systems such as the ones in \cite{Hangjun} and
\cite{Dukhyun} cannot be directly applied to the OFDMA systems.
For example, the scheme in \cite{Hangjun} requires iterative
ML estimations for all tones together with
clipping and filtering at the receiver, which is impractical for
the OFDMA systems. For OFDMA systems, the clipping noise
cancellation scheme is proposed in \cite{Ying}, but it is not easy
to be employed in the practical systems because the pilot tones
should be estimated at the receiver but they are distorted by
clipping.

In this paper, we propose a new clipping noise cancellation method
without data rate loss by using CS. The proposed scheme exploits
compressed observations of the clipping noise underlying in the
partial data tones to reconstruct the clipping noise without the
reserved tones. In the proposed scheme, the number of compressed
observations is adjustable in the sense that the optimal number of
compressed observations can be selected according to the noise
level. By doing so, we successfully overcomes the weakness of CS
reconstruction against noise. Numerical analysis shows that the
proposed method can mitigate the clipping noise well in both cases
of clipping at Nyquist sampling rate and clipping and filtering at
oversampling rate. Furthermore, the proposed scheme can also be
applied easily to OFDMA systems with interleaved partitioning.

This paper is organized as follows. The structure of OFDM system
and clipping scheme are reviewed in Section II. In Section III,
the proposed clipping noise cancellation scheme is presented. In
Section IV, the application of the proposed algorithm to OFDMA
systems is presented. Section V presents numerical analysis and
the conclusion is given in Section VI.

\section{System Model}
Let $X=(X(0),X(1),...,X(N-1))^T$ be an input symbol sequence.
The continuous time baseband OFDM signal can be represented as
\begin{equation}\label{eq:OFDMsig}
x(t)=\frac{1}{\sqrt{N}}\sum_{k=0}^{N-1}X(k)\exp \Bigg(\frac{j2\pi k t}{T}\Bigg),~~~~~0\leq t\leq T
\end{equation}
where $N$ is the number of subcarriers and $T$ is the symbol duration. Let $\Delta t_L=T/LN$ be a sampling interval,
where $L$ is oversampling factor. Then the discrete time OFDM signal sampled at time $n\Delta t_L$ can be expressed as
\begin{equation}
x_L(n)=x(n\Delta t_L),~~~~~n=0,1,...,LN-1.
\end{equation}
An OFDM signal sequence with oversampling factor $L$ can also be obtained by padding $X$ with $(L-1)N$ zeros and processing the inverse discrete Fourier transform (IDFT).

The PAPR of the OFDM signal sequence $x_L(n)$ with oversampling factor $L$ is defined as the ratio of the peak-to-average power of the signal as
\begin{equation}
\mathrm{PAPR}=\frac{\max_{0\leq n\leq LN-1} |x_L(n)|^2}{E[ |x_L(n)|^2 ]}
\end{equation}
where $E[\cdot]$ denotes the expected value.

To reduce peak regrowth at the D/A conversion, clipping and filtering are performed on the OFDM signal sequence oversampled by a factor $L$ at the transmitter. The clipping operation is given as
\begin{equation}
\bar{x}_L(n) =\begin{cases} x_L(n),                  & |x_L(n)| \leq A \\
                            A\exp \{j \mathrm{arg}(x_L(n)) \}, & |x_L(n)| > A
              \end{cases}
\end{equation}
where $A$ is the clipping threshold. Then the clipping ratio (CR) is defined as
\begin{equation}
\mathrm{CR} = 20 \log \frac{A}{\sigma}~[\mathrm{dB}]
\end{equation}
where $\sigma = \sqrt{E[|x_L(n)|^2]}$.

It is shown in \cite{Ochiai} that the clipped signal
$\bar{x}_L(n)$ can be modeled in two different ways. The first one
is an additive model where the clipped signal $\bar{x}_L(n)$ is
considered as the sum of the sampled signal $x_L(n)$ and the
clipping noise $c_L(n)$, and the second one is an attenuated
model where $\bar{x}_L(n)$ is viewed as the sum of an attenuated
component $\alpha x_L(n)$ and clipping noise component $d_L(n)$,
that is,
\begin{equation}
\bar{x}_L(n) =\begin{cases} x_L(n)+c_L(n),        & \mathrm{{for~additive~model}} \\
                            \alpha x_L(n)+d_L(n), & \mathrm{{for~attenuated~model}}
             \end{cases}
\end{equation}
where $n=0,1,...,LN-1$ and the attenuation $\alpha$ is given as \cite{Ochiai}
\begin{equation}
\alpha=1-e^{-\gamma^2}+\frac{\sqrt{\pi}\gamma}{2}\mathrm{erfc}(\gamma)
\end{equation}
and $\gamma=\sqrt{10^{\mathrm{CR}/10}}$.

In order to remove the out-of-band radiation due to clipping
operation, the clipped signal $\bar{x}_L(n)$ in time domain is
converted back to the frequency domain by taking $LN$-point
discrete Fourier transform (DFT) and being filtered. Then we can
have $\bar{X}(k)$ as
\begin{equation}\label{eq:barX}
\bar{X}(k) =\begin{cases} X(k)+C(k),              & \mathrm{{for~additive~model}} \\
                          \alpha X(k)+D(k),       & \mathrm{{for~attenuated~model}}
            \end{cases}
\end{equation}
where $k=0,1,...,N-1$. In \cite{Ochiai}, $D(k)$ is assumed as a
complex Gaussian random variable with zero mean.

Let $x(n)+c(n),~n=0,...,N-1,$ denote the IDFT of (\ref{eq:barX}) for the additive model, where $x(n)$ is Nyquist sampled OFDM
signal and $c(n)$ is clipping noise. For the case of $L=1$, we do
not use filtering and $c(n)$ is equal to $c_1(n)$. Then the
clipping noise $c(n)$ can be considered as $K$-sparse signal
having $K$ nonzero elements.
We define the sparsity ratio as $K/N$. Now we can exploit this
sparsity by CS method. If $L$ is larger than $1$, $c(n)$ is not
equal to $c_L(n)$ but $c(n)$ can be considered as nearly
$K$-sparse signal, implying that the amplitudes of remaining $N-K$
elements are close to zero, but not exactly zero. Even in this
case, CS method to cancel the clipping noise can also be applied
effectively.

\section{Clipping Noise Cancellation for OFDM Signals Using CS}

\subsection{Formulation to CS Problem}
The received symbol using the additive model in (\ref{eq:barX})
can be expressed in frequency domain as
\begin{equation}
Y(k) = H(k) (X(k) + C(k)) + Z(k),~~~~~0\leq k\leq N-1
\end{equation}
where $X(k)$ and $Y(k)$ denote the input and received symbols,
$H(k)$ denotes the frequency domain channel response, and $Z(k)$ denotes
the additive white Gaussian noise (AWGN) with variance $N_0$ in frequency domain.
In matrix form, it can be rewritten as
\begin{equation}
Y = \mathbf{H} (X+C) + Z
\end{equation}
where $\mathbf{H} = diag(H)$ and $Y$, $X$, $C$, $Z$ are $N\times 1$ column vectors.

We assume perfectly known channel response $\mathbf{H}$ and perfect synchronization. After channel equalization, its output is given as
\begin{equation}\label{eq:clippedmodel}
Y^{(eq)} = \mathbf{H}^{-1}Y =  X + C + \mathbf{H}^{-1}Z.
\end{equation}
Note that in the attenuated model, $Y^{(eq)}$ can be similarly
expressed as
\begin{equation}\label{eq:attnmodel}
Y^{(eq)} = \alpha X + D + \mathbf{H}^{-1}Z
\end{equation}
where the clipping noise $D$ is $N\times 1$ column vector.

\begin{figure}[t]
\centering
\includegraphics[width=18cm]{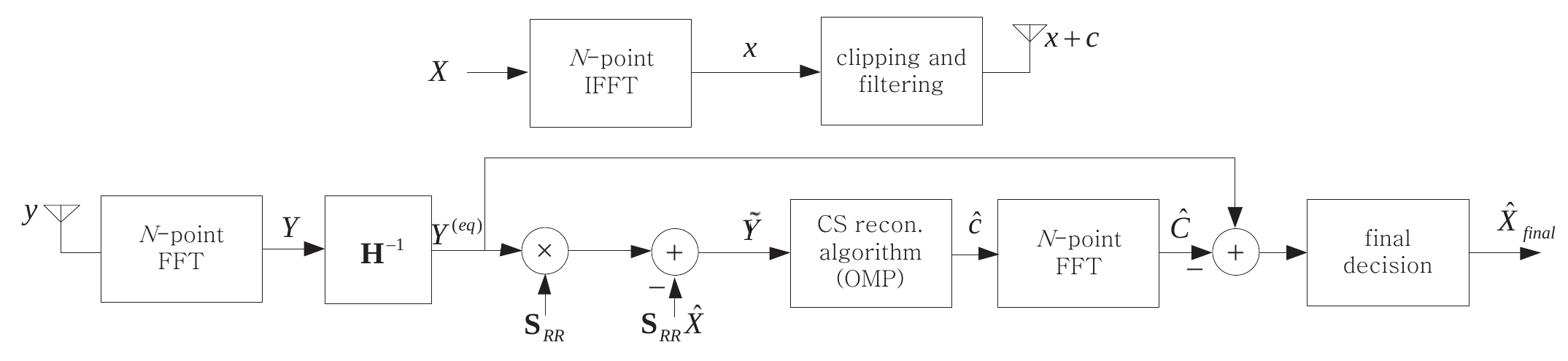}
\caption{Block diagram of the proposed scheme.}
\label{fig:block}
\end{figure}

For applying CS method, we need a compressed observation vector in the reduced
dimension. Our suggestion here is to select a subset of components
in $Y^{(eq)}$, namely $M$ out of $N$ components of $Y^{(eq)}$.
This can be done by multiplying $M\times N$ selection matrix
$\mathbf{S}_{RR}$ consisting of some $M$ rows of the identity
matrix $\mathbf{I}_N$ to $Y^{(eq)}$ in (\ref{eq:clippedmodel}). The
discussion on $\mathbf{S}_{RR}$ will be made in the next
subsection.
Let $C=\mathbf{F}c$, where
$\mathbf{F}$ is $N\times N$ DFT matrix. Then, we have
\begin{equation}\label{eq:SYeq}
\mathbf{S}_{RR} Y^{(eq)} = \mathbf{S}_{RR} \mathbf{F}c + \mathbf{S}_{RR} X + \mathbf{S}_{RR} \mathbf{H}^{-1}Z.
\end{equation}
If we subtract the estimation $\mathbf{S}_{RR} \hat{X}$ from
(\ref{eq:SYeq}), we have
\begin{align}\label{eq:CSform}
\tilde{Y} = \mathbf{S}_{RR}Y^{(eq)}-\mathbf{S}_{RR}\hat{X} &= \mathbf{S}_{RR} \mathbf{F}c + \mathbf{S}_{RR} (X-\hat{X}) + \mathbf{S}_{RR} \mathbf{H}^{-1}Z \nonumber\\
   &= \mathbf{\Phi} c + \underbrace{\mathbf{S}_{RR} (X-\hat{X}) + \mathbf{S}_{RR} \mathbf{H}^{-1}Z}_{\mathrm{noise~vector}}
\end{align}
where the matrix $\mathbf{\Phi} = \mathbf{S}_{RR} \mathbf{F}$ can
be considered as $M\times N$ \textit{measurement matrix} in CS. As one
can see in \cite{Candes}, the measurement matrix for CS can be
constructed by using the subset of rows in DFT matrix. Then the
resulting equation (\ref{eq:CSform}) can be considered as CS
problem, where the vector $\tilde{Y}$ can be considered as $M\times 1$
\textit{compressed observation vector}, the clipping noise $c$ as
$N\times 1$ \textit{sparse signal vector}, and the remaining
vector $\mathbf{S}_{RR} (X-\hat{X}) + \mathbf{S}_{RR}
\mathbf{H}^{-1}Z$ as $M\times 1$ \textit{noise vector}. By using
CS reconstruction algorithm, we can reconstruct $c$ as $\hat{c}$
from the compressed observation vector $\tilde{Y}$. Then fast Fourier
transform (FFT) of $\hat{c}$ is subtracted from the equalized
received symbol $Y^{(eq)}$ and then the final decision is made.
Fig. \ref{fig:block} pictorially summarizes the proposed scheme.

\subsection{Selection Matrix $\mathbf{S}_{RR}$}
\begin{figure}[htbp]
\centering
\includegraphics[width=15cm]{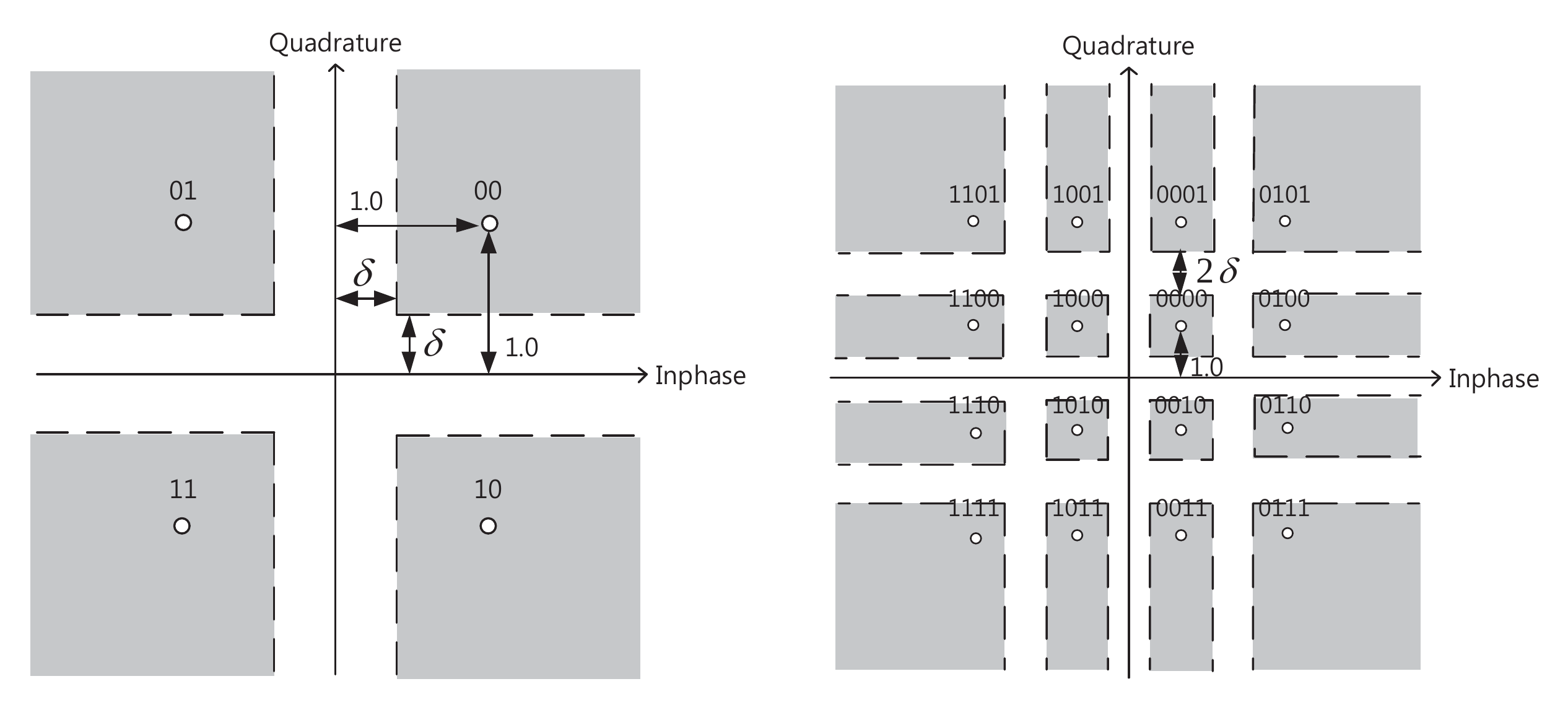}
\caption{Reliable regions for QPSK and 16-QAM modulations.}
\label{fig:RR}
\end{figure}

As mentioned in the previous subsection, we suggest that the
$M\times 1$ compressed observation vector $\tilde{Y} = \mathbf{S}_{RR}Y^{(eq)}-\mathbf{S}_{RR}\hat{X}$. And, $\mathbf{S}_{RR}Y^{(eq)}$ in $\tilde{Y}$ is constructed from $Y^{(eq)}$ by
selecting those components of $Y^{(eq)}$ that are thought to be
more reliable than others. We designate a specific area in the
received signal space as reliable region (RR) and define the index
set $\mathcal{K}_{RR}$ as the set of component indices of the
received signals that lie in the reliable region. For quaternary
phase shift keying (QPSK) and 16-quadrature amplitude modulation
(QAM), RR is designated as shaded area in Fig. \ref{fig:RR}. For
other modulations, RR can be set in analogous way. In Fig.
\ref{fig:RR}, the distance between adjacent signal points is
assumed to be $2$.

Prior to explain the index set $\mathcal{K}_{RR}$ in detail, we compare the additive model and the attenuated model in (\ref{eq:barX}) from the viewpoint of signal-to-clipping-noise ratio (SCNR).
In \cite{Ochiai}, $D(k)$ in attenuated model is uncorrelated clipping noise
from $X(k)$. And, $C(k)$ in additive model is easily derived from (\ref{eq:barX}) as
\begin{equation}\label{eq:CandD}
C(k) = (\alpha-1)X(k) + D(k).
\end{equation}
Additionally to $D(k)$, $C(k)$ in (\ref{eq:CandD}) also has a term $(\alpha-1)X(k)$ correlated from
$X(k)$. And thus it implies that the
SCNR of the attenuated model is
larger than SCNR of the additive model, that is,
\begin{equation}\label{eq:SCNR}
\frac{E[|\alpha X(k)|^2]}{E[|D(k)|^2]} \geq
\frac{E[|X(k)|^2]}{E[|C(k)|^2]}.
\end{equation}

For the reason as above, we describe the index set $\mathcal{K}_{RR}$ in detail using the attenuated model.
Assume that $M$ components of $(1/\alpha) Y^{(eq)}$ fall into RR. Then the index set $\mathcal{K}_{RR}$ can be expressed as
\begin{equation}\label{eq:RR}
\mathcal{K}_{RR}=\{ k_m:\frac{1}{\alpha}Y^{(eq)}(k_m)\in \mathrm{RR},~0\leq m\leq M-1 \}.
\end{equation}
The $M\times N$ selection matrix $\mathbf{S}_{RR}$ can be obtained
from the identity matrix of order $N$ by selecting $M$ rows
corresponding to the index set $\mathcal{K}_{RR}$, where $M$ is
cardinality of the set $\mathcal{K}_{RR}$.

In our proposed scheme, for the determined $\mathbf{S}_{RR}$, we estimate $\mathbf{S}_{RR} X$. Clearly, we can also conclude that $\mathbf{S}_{RR} X$ can be more
efficiently estimated by using attenuated model. That is, in order to estimate $\mathbf{S}_{RR} X$ efficiently, we decide $(1/ \alpha)\mathbf{S}_{RR} Y^{(eq)}$ as $\mathbf{S}_{RR} \hat{X}$ by ML estimation. Intuitionally, at the receiver, when we
decide $(1/\alpha) Y^{(eq)}(k_m)$ in RR as $\hat{X}(k_m)$, its decision error probability is lower
than that when we decide $(1/\alpha) Y^{(eq)}(k \not \in \mathcal{K}_{RR})$ in the outside region of RR as $\hat{X}(k \not \in \mathcal{K}_{RR})$. Therefore, if we consider
the signals only in the RR to reconstruct the clipping noise $c$,
decision error of $\mathbf{S}_{RR} \hat{X}$ can be reduced as $\delta$ increases in Fig. \ref{fig:RR}.

\subsection{Decision Error Probability of $\mathbf{S}_{RR} \hat{X}$ Over an AWGN Channel}

For an AWGN channel, after scaling by $1/\alpha$,
(\ref{eq:attnmodel}) can be rewritten as
\begin{equation}\label{eq:awgn}
\frac{1}{\alpha} Y(k) = \frac{1}{\alpha}Y^{(eq)}(k) =  X(k) + \frac{1}{\alpha}W(k),~~~k=0,1,...,N-1
\end{equation}
where $W(k)=D(k)+Z(k)$. Since $D(k)$ is assumed to be complex
Gaussian random variable with zero mean, $(1/\alpha)W(k)$ in
(\ref{eq:awgn}) can be characterized as a zero mean complex
Gaussian random variable with variance $E[|(1/\alpha)
W(k)|^2]=(1/\alpha^2)(E[|D(k)|^2]+N_0)$, statistically independent
of the input symbol \cite{Ochiai}. We only consider the case of
clipping at Nyquist sampling rate. For clipping at Nyquist
sampling rate,
all the data tones experience the
clipping noise $D(k)$ with the same variance as
\begin{equation}
E[|D(k)|^2]=(1-e^{-\gamma ^2}- \alpha ^2)E[|X(k)|^2].
\end{equation}
And thus the variance of $(1/\alpha) W(k)$, denoted by $N_0' = E[
|(1/\alpha) W(k)|^2 ]$ is the same for each data tone. The
decision error probability of $\mathbf{S}_{RR} \hat{X}$ for $V^2$-QAM of
even $V$ is expressed as
\begin{align}\label{eq:proba}
P_{V^2}\Big(\hat{X}(k) \neq X(k)~|~\frac{1}{\alpha} Y(k) \in RR \Big)
&= 1-P_{V^2} \Big(\hat{X}(k) = X(k)~|~\frac{1}{\alpha} Y(k) \in RR \Big)\nonumber\\
&= 1-\frac{P_{V^2}(\hat{X}(k) = X(k) \cap \frac{1}{\alpha} Y(k) \in RR)}{P_{V^2}(\frac{1}{\alpha} Y(k) \in RR)}.
\end{align}

In (\ref{eq:proba}), $P_{V^2}((1/\alpha) Y(k) \in RR)$, the
probability that $(1/\alpha)Y(k)$ belongs to RR for $V^2$-QAM is
rewritten as
\begin{equation}
P_{V^2}\Big(\frac{1}{\alpha} Y(k) \in RR \Big) = \Big[ P_{V} \Big(\frac{1}{\alpha} Y(k) \in RR \Big) \Big]^2
\end{equation}
where $P_{V}((1/\alpha) Y(k) \in RR)$ is the probability that
inphase (or quadrature) component of $(1/\alpha)Y(k)$ belongs to
RR for $V$-pulse amplitude modulation (PAM).

Then, $P_{V}((1/\alpha) Y(k) \in RR)$ is expressed as
\begin{align}\label{eq:YinRR}
P_{V} \Big(\frac{1}{\alpha} Y(k) \in RR \Big)
=&\frac{2}{V}\sum_{v=0}^{(V-2)/2} \Bigg(
\int_{-\infty}^{2-V-\delta} \frac{1}{\sqrt{\pi N_0'}}e^{-(x-(2v+1))^2/N_0'}~dx \nonumber\\
&+ \int_{2-V+\delta}^{4-V-\delta} \frac{1}{\sqrt{\pi N_0'}}e^{-(x-(2v+1))^2/N_0'}~dx
+ \int_{4-V+\delta}^{6-V-\delta} \frac{1}{\sqrt{\pi N_0'}}e^{-(x-(2v+1))^2/N_0'}~dx  \nonumber\\
&+ ...+ \int_{-4+V+\delta}^{-2+V-\delta} \frac{1}{\sqrt{\pi N_0'}}e^{-(x-(2v+1))^2/N_0'}~dx
+ \int_{-2+V+\delta}^{\infty} \frac{1}{\sqrt{\pi N_0'}}e^{-(x-(2v+1))^2/N_0'}~dx
\Bigg) \nonumber\\
=&\frac{2}{V}\sum_{v=0}^{(V-2)/2}
\Bigg\{
Q\Bigg( \frac{2v-1+V+\delta}{\sqrt{N_0'/2}} \Bigg)
+ Q\Bigg( \frac{-2v-3+V+\delta}{\sqrt{N_0'/2}} \Bigg)\nonumber\\
&+\sum_{v'=0}^{V-3}
\Bigg(
Q\Bigg( \frac{-2v+2v'+1-V+\delta}{\sqrt{N_0'/2}} \Bigg)
- Q\Bigg( \frac{-2v+2v'+3-V-\delta}{\sqrt{N_0'/2}} \Bigg)
\Bigg)
\Bigg\}
\end{align}
where $Q(z)=\int_{z}^{\infty} \frac{1}{\sqrt{2\pi}}e^{-y^2/2}~dy$.

The probability $P_{V^2}(\hat{X}(k) = X(k) \cap (1/\alpha) Y(k)
\in RR)$ in (\ref{eq:proba}) is equivalent to the correct decision probability for $V^2$-QAM when
distance between adjacent signal points is $2-2\delta$ and it is
given in \cite{Proakis} as
\begin{equation}\label{eq:YinRRandcrt}
P_{V^2} \Big(\hat{X}(k) = X(k) \cap \frac{1}{\alpha} Y(k) \in RR \Big)
=\Bigg\{ 1-\frac{2(V-1)}{V}Q \Bigg( \frac{1-\delta}{\sqrt{N_0'/2}} \Bigg)\Bigg\}^2.
\end{equation}
Consequently, the decision error probability of $\mathbf{S}_{RR} \hat{X}$ is obtained by substituting (\ref{eq:YinRR}) and (\ref{eq:YinRRandcrt}) into (\ref{eq:proba}).

\subsection{Determination of $\delta$}

Accuracy of reconstruction via CS in (\ref{eq:CSform}) is affected
by not only the noise but also the selection of compressed
observations. Certainly, both the noise and the compressed
observations vary according to $\delta$ in our scheme.
The influence of $\delta$ on the performance of CS reconstruction
can be explained in two different views.

Firstly, the amount of estimation errors is dependent on the value
of $\delta$. If $\delta$ is set to be too small, then estimation
error will occur frequently, which results in increasing the
energy of $|\mathbf{S}_{RR} (X - \hat{X})|$ in the noise vector in
(\ref{eq:CSform}). Accordingly, the performance of CS
reconstruction will be degraded and thus we can say that large
value of $\delta$ is preferable in this point of view.

Secondly, the number of compressed observations which is the row
size $M$ of $\mathbf{\Phi} = \mathbf{S}_{RR}\mathbf{F}$ is
dependent on $\delta$. As $\delta$ decreases, the RR becomes
wider, which implies that $M$ becomes larger as
\begin{equation}\label{eq:avgM}
E[M] = N\cdot P_{V^2} \Big(  \frac{1}{\alpha} Y(k) \in RR \Big).
\end{equation}
In our proposed scheme, for clipping at Nyquist sampling rate over an AWGN
channel, the probability that the component $k$ is in
$\mathcal{K}_{RR}$
is identical for all $k$. Accordingly, $\mathbf{\Phi}$ can be
considered as random selection of $M$ rows from $N$-point DFT
matrix. In \cite{Candes2}, for randomly selected compressed observations in frequency domain, the quality of the reconstruction is upper bounded as
\begin{equation}
\parallel c- \hat{c} \parallel _2 \leq O(1)\cdot R \cdot (M/(\log{N})^6)^{-r},
\end{equation}
where $r$ and $R$ are related to the sparsity of the clipping noise $c$. And thus, for fixed $N$ and $c$, large $M$ improves the reconstruction performance.
Therefore, in our proposed scheme, small value of $\delta$ is
preferred in this point of view.

Consequently, we have to consider above two aspects for
determining $\delta$. We perform the massive simulations to find
an optimal value of $\delta$ which maximizes the performance of CS
reconstruction.

\subsection{Reconstruction Algorithms in CS}

A number of reconstruction algorithms have been introduced in many literature of the CS \cite{Donoho}-\cite{Candes2}.
We adopt a sparse approximation algorithm called orthogonal
matching pursuit (OMP) to handle the signal recovery problem
\cite{Tropp}. The main reason of selecting OMP algorithm is not
its performance but its ease of implementation and
speed. Other reconstruction algorithms such as basis pursuit (BP) can show better
reconstruction performance, while they require high computational
complexity.

Since there is no information about sparsity $K$ of clipping noise
$c$ at the receiver, we need to set the iteration number of OMP
sufficiently large.
As OMP algorithm runs iteratively, it estimates the nonzero
clipping noise in descending order of amplitude. The clipping
noise having small amplitude below a certain threshold can be
negligible. Therefore, at each iteration of OMP, when the recent
reconstructed clipping noise component is below the properly predetermined
threshold or the maximum number of iterations is reached, we can
stop the OMP algorithm to reduce the unnecessary computational
complexity.

In this paper, we investigate into the case of
clipping and filtering ($L > 1$) and the case of clipping at
Nyquist sampling rate ($L=1$). Additionally,
the $K$ largest peaks reduction scheme similar to the peak
annihilation in \cite{Al-Safadi} is also simulated. That is, we
scale down exactly $K$ components having large amplitudes in the
OFDM signal sequence $x$ at Nyquist sampling rate. These
components are scaled down to have $80\%$ of $(K+1)$-th largest OFDM
signal's amplitude to mitigate peak regrowth after D/A conversion.
In this case, the OMP needs to be done $K$ times.


\section{Clipping Noise Cancellation for Interleaved OFDMA Downlink Systems}

Different from OFDM systems, in the downlink of OFDMA systems,
each user only has knowledge of its own data tones. Moreover,
users in OFDMA system can use different modulation schemes in
general. Since it is not always possible for a particular user to
demodulate the data tones of the other users, the estimation
$\mathbf{S}_{RR} \hat{X}$ of the data tones in the RR could be
impossible at the receiver. Hence the proposed scheme needs to be
modified in order to be applied to OFDMA systems.

In a $2^U$-user OFDMA downlink system with $N = 2^b$ subcarriers,
all the available subcarriers can be partitioned into $2^U$
interleaved subcarrier groups. For each subcarrier group assigned
to each user, small FFT block can be extracted from the $N$ point
FFT block. The main idea of this section is exploiting this
structure.

\begin{figure}[htbp]
\centering
\includegraphics[width=17cm]{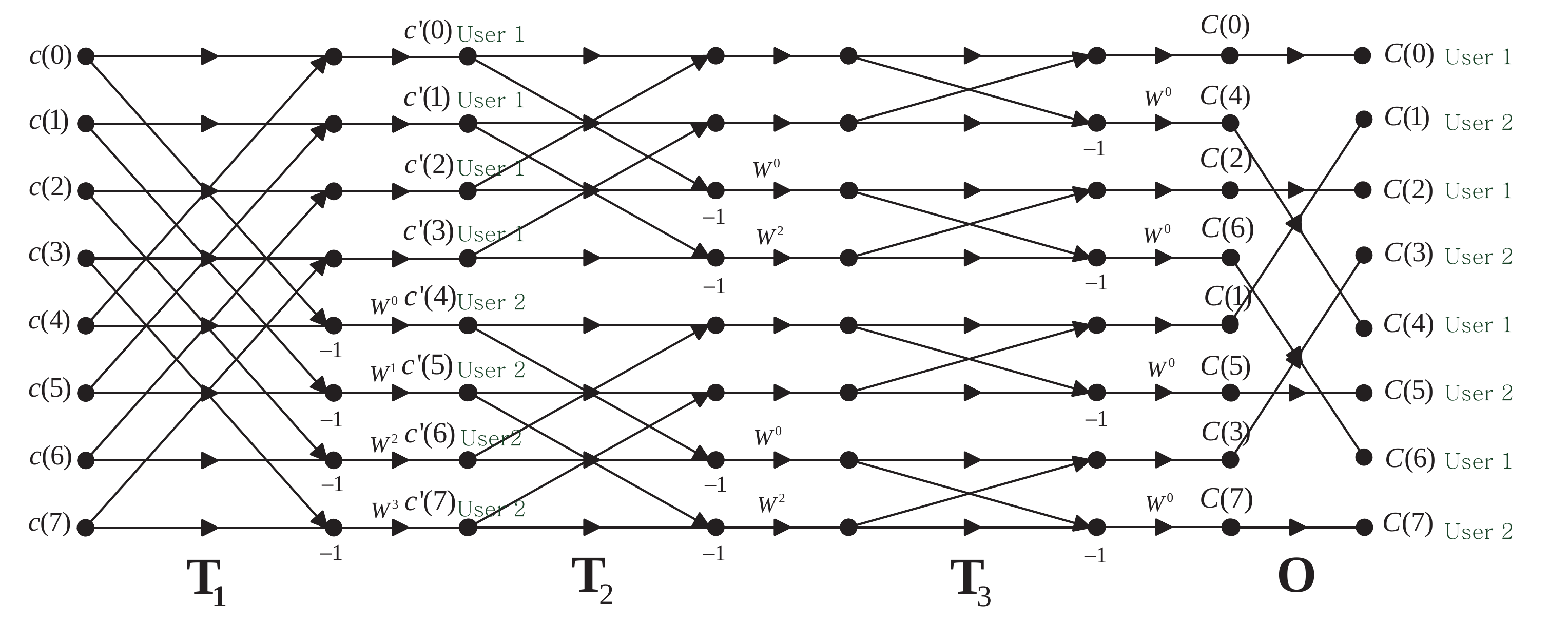}
\caption{Decimation-in-frequency FFT when $N=8~(b=3)$ and $U=1$.}
\label{fig:OFDMA}
\end{figure}

Fig. \ref{fig:OFDMA} shows the structure of
decimation-in-frequency FFT of the clipping noise $c$ when $N=8$
and $U=1$. As one can see in Fig. \ref{fig:OFDMA}, subcarriers are
assigned to user 1 and user 2 by interleaved pattern. In general,
$N$-point FFT consists of $b$ stages and then this can also be
represented by the matrix form as
\begin{equation}\label{eq:decomFFT}
\mathbf{F}=\mathbf{O} \mathbf{T}_b \mathbf{T}_{b-1}...\mathbf{T}_2 \mathbf{T}_1
\end{equation}
where $\mathbf{T}_i,~1\leq i\leq b,$ denotes the $i$-th stage
matrix of FFT process and $\mathbf{O}$ is an index ordering
matrix. For instance, $8$-point FFT block can be represented by
the matrix form $\mathbf{OT}_3\mathbf{T}_2\mathbf{T}_1$ as shown
in Fig. \ref{fig:OFDMA}. Similarly, for $N/2^U$-point FFT, we also
define the matrices $\mathbf{O}'$ and $\mathbf{T}'_{i'},~1\leq i'\leq
b-U,$ as the index ordering matrix and $i'$-th stage matrix of FFT
process, respectively.

Let $\mathbf{I}_N$ be the identity matrix of order $N$
and $e_j$ be the $j$-th row of $\mathbf{I}_N$. Let
$\mathbf{P}^I_u$ and $\mathbf{P}^A_u$ be $N/2^U\times N$
interleaved and adjacent partitioning matrices for the $u$-th user
defined as $\mathbf{P}^I_u = [e^T_u e^T_{2^U+u} ...
e^T_{N-2^U+u}]^T$ and $\mathbf{P}^A_u=(\mathbf{O}')^{-1}
\mathbf{P}^I_u \mathbf{O}$. For instance, in Fig. \ref{fig:OFDMA}, for user 1, $\mathbf{P}^I_1$
selects the subset of components in the clipping noise $C$ (after $\mathbf{O}$) denoted as ``user 1'' (i.e., $C(0)$,~$C(2)$,~$C(4)$,~$C(6)$). And, $\mathbf{P}^A_1$ selects the subset of components between $\mathbf{T}_1$ and $\mathbf{T}_2$ denoted
as ``user 1'' (i.e., $c'(0)$, $c'(1)$, $c'(2)$, $c'(3)$).

In order to extract the $N/2^U$-point FFT block for the $u$-th user from the $N$-point FFT block for OFDMA signals, we multiply (\ref{eq:clippedmodel}) by the $u$-th user interleaved partitioning matrix $\mathbf{P}^I_u$. And using (\ref{eq:decomFFT}), (\ref{eq:clippedmodel}) can be modified as
\begin{align}\label{eq:OFDMAmodel}
\mathbf{P}^I_u Y^{(eq)}
&= \mathbf{P}^I_u X + \mathbf{P}^I_u (\mathbf{O} \mathbf{T}_b...\mathbf{T}_1) c + \mathbf{P}^I_u \mathbf{H}^{-1}Z \nonumber\\
&= \mathbf{P}^I_u X + \mathbf{P}^I_u (\mathbf{O} \mathbf{T}_b...\mathbf{T}_{U+1}) (\mathbf{T}_U ... \mathbf{T}_1 c) + \mathbf{P}^I_u \mathbf{H}^{-1}Z.
\end{align}

Clearly, $\mathbf{P}^I_u\mathbf{O}$ can be modified as $\mathbf{O}'\mathbf{P}^A_u$ and it is easy to check that $\mathbf{P}^A_u(\mathbf{T}_b...\mathbf{T}_{U+1})=(\mathbf{T}'_{b-U}...\mathbf{T}'_1)\mathbf{P}^A_u$.
Then (\ref{eq:OFDMAmodel}) can be rewritten as
\begin{align}\label{eq:OFDMAmodel2}
\mathbf{P}^I_u Y^{(eq)}
&= \mathbf{P}^I_u X + \mathbf{O}'\mathbf{P}^A_u(\mathbf{T}_b...\mathbf{T}_{U+1}) (\mathbf{T}_U...\mathbf{T}_1c)  + \mathbf{P}^I_u \mathbf{H}^{-1}Z \nonumber\\
&= \mathbf{P}^I_u X + \mathbf{O}'(\mathbf{T}'_{b-U}...\mathbf{T}'_1)\mathbf{P}^A_u (\mathbf{T}_U...\mathbf{T}_1c)  + \mathbf{P}^I_u \mathbf{H}^{-1}Z \nonumber\\
&= \mathbf{P}^I_u X + \mathbf{F}'c' + \mathbf{P}^I_u \mathbf{H}^{-1}Z
\end{align}
where $\mathbf{F}'=\mathbf{O}'(\mathbf{T}'_{b-U}...\mathbf{T}'_1)$ is $N/2^U \times N/2^U$ DFT matrix and $c'=\mathbf{P}^A_u \mathbf{T}_U...\mathbf{T}_1 c$.

Since $N/2^U\times N$ matrix $\mathbf{P}^A_u \mathbf{T}_U...\mathbf{T}_1$ has only one nonzero elements per a column, if $c$ is $K$ sparse, then $c'=\mathbf{P}^A_u \mathbf{T}_U...\mathbf{T}_1 c$ is also sparse signal having the maximum $K$ sparsity but the reduced dimension from $N$ to $N/2^U$. Therefore, the sparsity ratio of clipping noise $c'$ in the OFDMA system effectively increases compared to $c$ in the OFDM system in (\ref{eq:CSform}). And, it may somewhat degrade reconstruction performance.

Now, as we have previously done in (\ref{eq:SYeq}) and (\ref{eq:CSform}) for OFDM systems, we can formulate (\ref{eq:OFDMAmodel2}) to the CS problem as
\begin{align}\label{eq:CSformOFDMA}
\tilde{Y}'=
\mathbf{S}_{RR}' \mathbf{P}^I_u Y^{(eq)} - \mathbf{S}_{RR}' \mathbf{P}^I_u \hat{X}
&= \mathbf{S}_{RR}' \mathbf{F}'c' + \mathbf{S}_{RR}' \mathbf{P}^I_u (X-\hat{X}) + \mathbf{S}_{RR}'\mathbf{P}^I_u \mathbf{H}^{-1}Z \nonumber\\
&= \mathbf{\Phi}'c' + \underbrace{\mathbf{S}_{RR}' \mathbf{P}^I_u (X-\hat{X}) + \mathbf{S}_{RR}'\mathbf{P}^I_u \mathbf{H}^{-1}Z}_{\mathrm{noise~vector}}
\end{align}
where $\mathbf{S}_{RR}'$ and $\mathbf{\Phi}'=\mathbf{S}_{RR}' \mathbf{F}'$ are selection matrix and a measurement matrix for OFDMA systems, respectively. In (\ref{eq:CSformOFDMA}), it is possible to estimate $\mathbf{S}_{RR}' \mathbf{P}^I_u X$ because it requires only the data tones of $u$-th user. In the proposed noise cancellation scheme for OFDMA systems, from the compressed observation vector $\tilde{Y}'$, we reconstruct the clipping noise $c'$, which is associated with only the data tones of $u$-th user, instead of reconstructing the whole clipping noise $c$. For instance, in Fig. \ref{fig:OFDMA}, clipping noise $c'(0),c'(1),c'(2),c'(3)$ is reconstructed for user 1 instead of clipping noise $c(0),c(1),c(2),...,c(7)$.

\section{Numerical Analysis}

In this section, we evaluate the BER performance of the proposed
clipping noise cancellation scheme in the AWGN channel. We simulate
both the case of clipping at Nyquist sampling rate with $L=1$ and
the case of clipping and filtering at the oversampling rate with
$L=4$. Additionally, the $K$ largest peaks reduction scheme is also
simulated. The distance between adjacent signal points is $2$.

\begin{figure}[H]
\centering
\includegraphics[width=11cm]{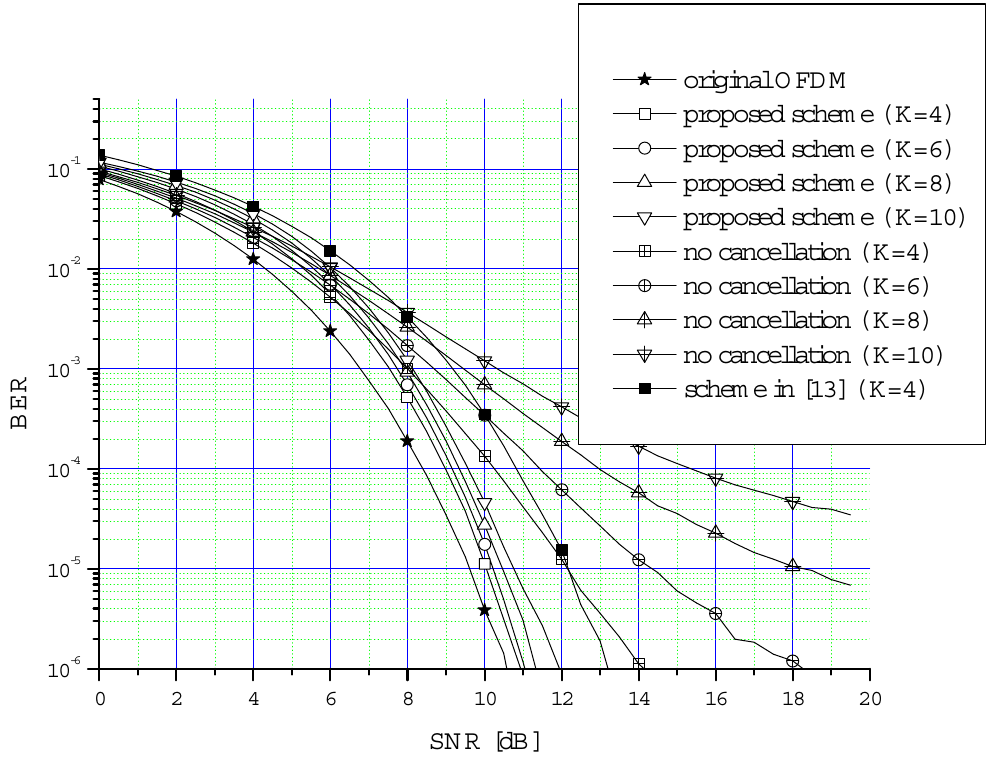}
\caption{BER performance of the proposed scheme and the scheme in \cite{Al-Safadi} with $K$ largest peaks reduction when $N=64$ and QPSK is used.}
\label{fig:scdown}
\end{figure}
Fig. \ref{fig:scdown} shows the BER performance of the proposed scheme using $K$ largest peaks reduction. There is a large benefit to use the proposed scheme compared to no clipping noise cancellation case. As $K$ increases, the BER performance decreases and for $K>8$, the BER performance is not closed to that of the original OFDM. Thus, we use the maximum iteration number of OMP as $0.125N$ for unknown $K$ cases. In Fig. \ref{fig:scdown}, clipping noise cancellation scheme in \cite{Al-Safadi} is also given when $29$ tones are positioned by a $(59,29,14)$ difference set \cite{diffset} and reserved. It uses the sufficient number of reserved tones (i.e., the sufficient number of compressed observations) for reconstructing the clipping noise with $K=4$ in the absence of noise. But, the scheme in \cite{Al-Safadi} shows the poor BER performance due to weakness of CS reconstruction against noise. Although the clipping noise cancellation scheme in \cite{Mohammad} is not shown in Fig. \ref{fig:scdown}, it may show worse BER performance than that in \cite{Al-Safadi}. The reason is that, considering the practical system, the number of pilot tones (i.e., the number of compressed observations) is smaller than $29$ when $N=64$.

\begin{figure}[H]
\centering
\includegraphics[width=11cm]{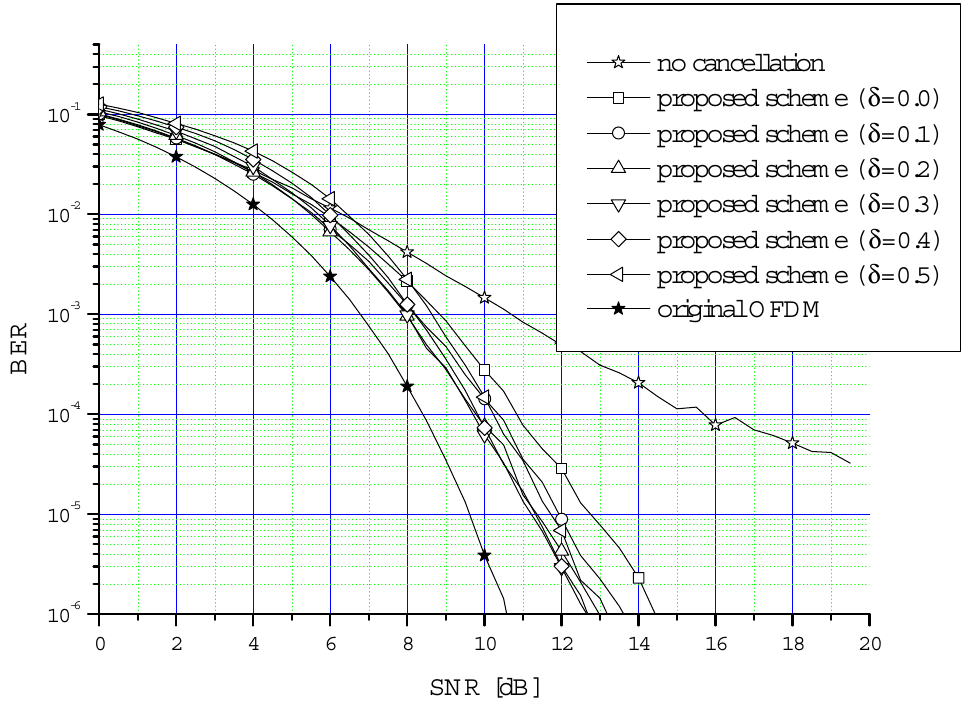}
\caption{BER performance of the proposed scheme with clipping at Nyquist sampling rate when CR $=0\mathrm{dB}$, $N=64$, and QPSK is used.}
\label{fig:Nyquist}
\end{figure}
Fig. \ref{fig:Nyquist} shows the BER performance of the proposed scheme using clipping at Nyquist sampling rate for various $\delta$ when $N=64$ and QPSK is used. We set the maximum iteration number for OMP as $8$. It shows that $\delta=0.4\sim 0.5$ gives the best BER performance. As we mentioned, smaller or larger value than $\delta=0.4\sim 0.5$ causes frequent decision error or lack of compressed observations. Thus it induces the degradation of BER performance. In Fig. \ref{fig:Nyquist}, since CR$=0\mathrm{dB}$ is small value, $K$ nonzero components of the clipping noise $c$ is larger than $8$. Therefore, OMP cannot reconstruct the remaining nonzero components in the clipping noise $c$ and BER performance is not closed to that of the original OFDM. Additionally, for $\delta=0.0$, we do not need to use CS reconstruction algorithm because $M=N$. Using largest $8$ components in IDFT of the compressed observation vector for reconstructed clipping noise $\hat{c}$ gives us almost the same BER performance as using CS reconstruction.

\begin{figure}[H]
\centering
\subfigure[]{
\includegraphics[width=11cm]{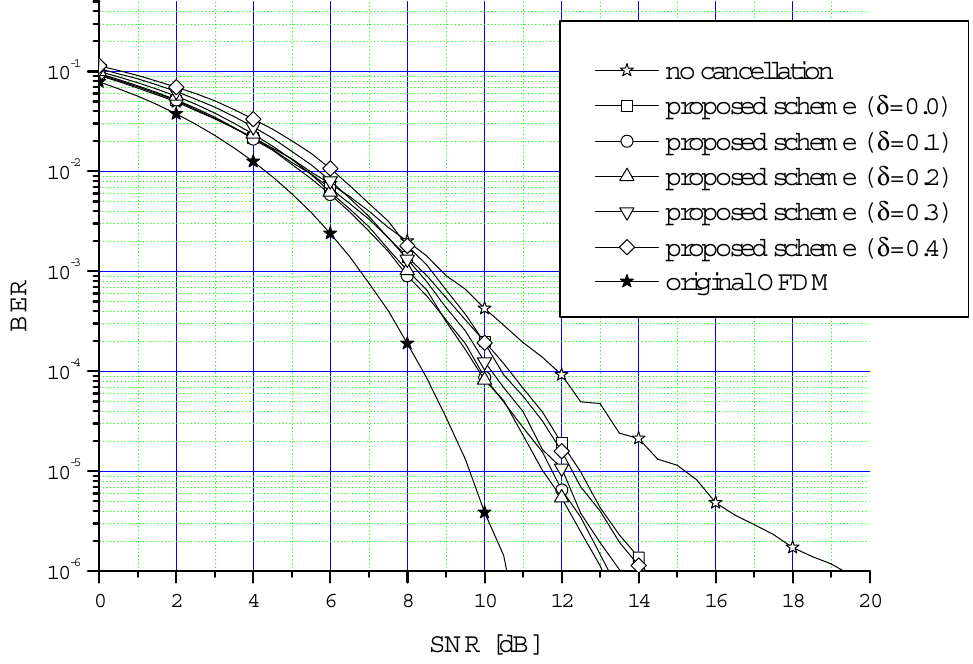}
}
\subfigure[]{
\includegraphics[width=11cm]{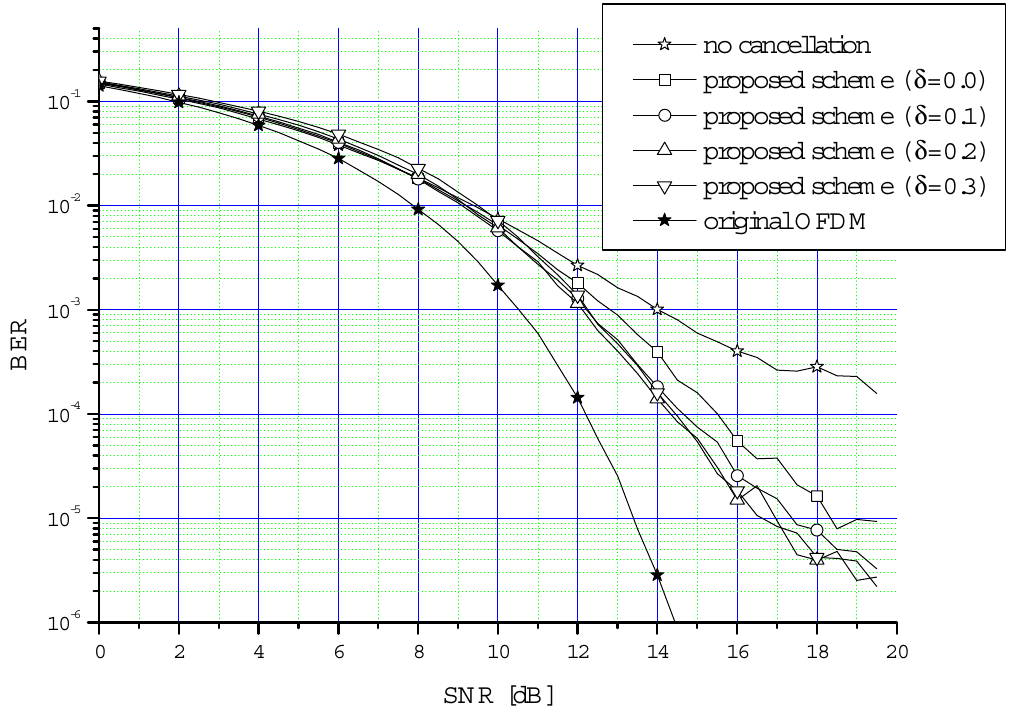}
}
\end{figure}
\begin{figure}[H]
\centering
\subfigure[]{
\includegraphics[width=11cm]{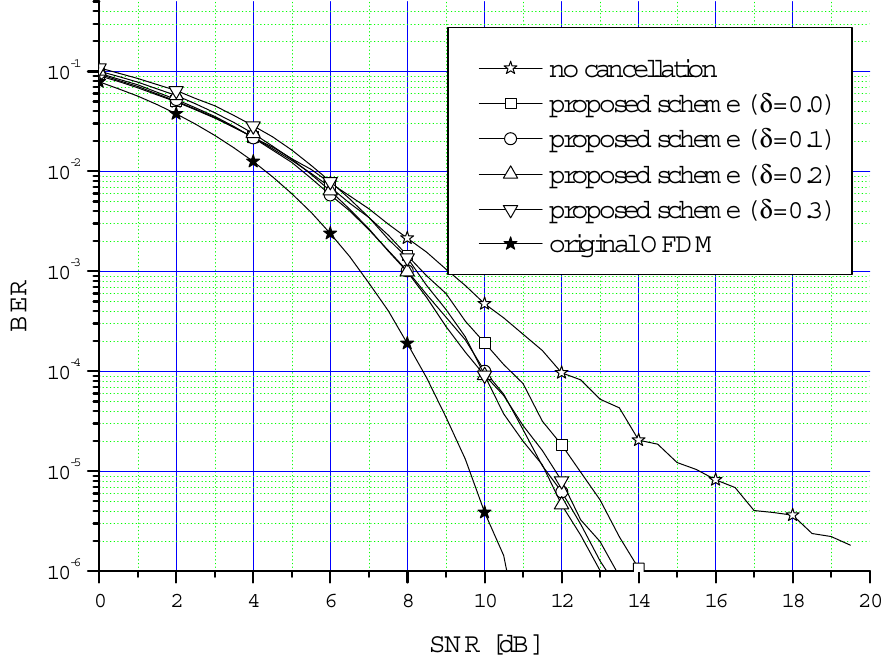}
}
\caption{BER performance of the proposed scheme with clipping and filtering ($L=4$) (a) CR $=0\mathrm{dB}$, $N=64$, and QPSK is used. (b) CR $=3\mathrm{dB}$, $N=64$, and 16-QAM is used. (c) CR $=0\mathrm{dB}$, $N=512$, and QPSK is used.}
\label{fig:CF}
\end{figure}
Fig. \ref{fig:CF} shows the BER performance when clipping and filtering are performed ($L=4$) and we reconstruct the clipping noise $c$ by using OMP for several $\delta$, $N$, and modulation schemes. We set the maximum iteration number for OMP as $0.125N$. Similar to Fig. \ref{fig:Nyquist}, there is a $\delta$ which can achieve the best BER performance. Since clipping noise $c$ for $L>1$ has nearly sparsity, it shows the limited performance compared to clipping at Nyquist sampling rate. Fig. \ref{fig:CF} also shows the BER performance of the proposed scheme for large $N$. For large $N$, the BER performance is almost the same as that of small $N$. And, $\delta$ for the best BER performance is the same. Therefore, we can conclude that $\delta$ for the best BER performance is mainly affected by CR and modulation order.

\begin{figure}[H]
\centering
\includegraphics[width=11cm]{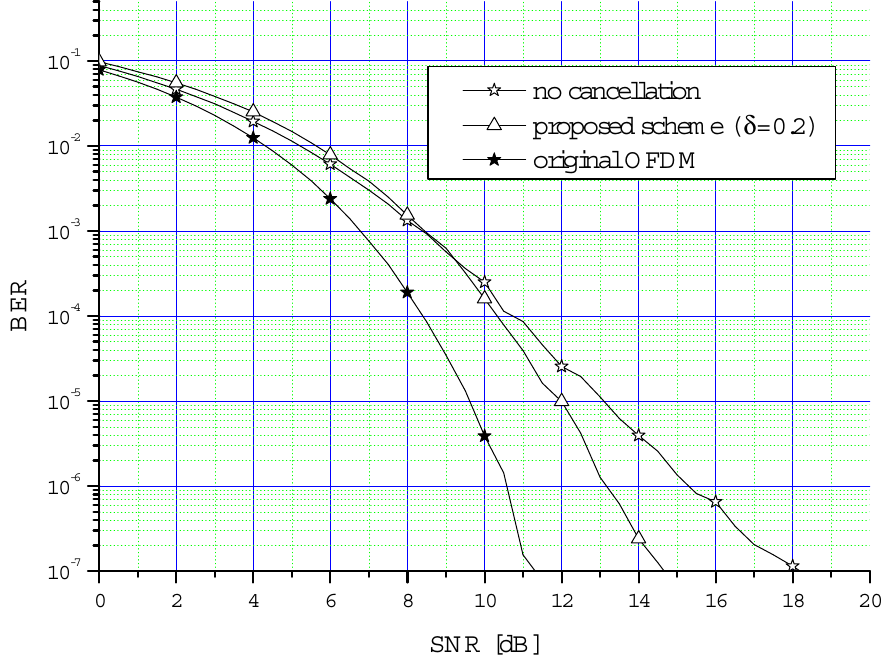}
\caption{BER performance of the proposed scheme applied to OFDMA system for $4$ users with clipping at Nyquist sampling rate when CR $=2\mathrm{dB}$, $N=512$, and QPSK is used.}
\label{fig:BEROFDMA}
\end{figure}
Fig. \ref{fig:BEROFDMA} shows the BER performance when the proposed scheme is applied to interleaved OFDMA systems. Clipping at Nyquist sampling rate is used for $N=512$ and $2^U=4$ users. And we set the maximum iteration number for OMP as $16$ and $\delta = 0.2$ is used. We consider only the $128$ data tones corresponding to user 1 and also construct only the clipping noise for user 1. Therefore, its computational complexity is reduced compared to OFDM system for $N=512$ because the size of measurement matrix decreased. However, sparsity ratio increases by the maximum $4$ times and it causes poor reconstruction performance.

\section{Conclusion}
In this paper, we proposed the new clipping noise cancellation scheme in OFDM using CS. In the proposed scheme, the data tones are partially exploited to reconstruct the clipping noise instead of the whole data tones. By introducing RR, we can select these partial data tones and numerical analysis shows that the proper value of $\delta$ needs to be chosen. Additionally, the proposed scheme  can also be applied easily to OFDMA systems because of the FFT structure. Using the proposed clipping noise cancellation scheme for the OFDM and OFDMA systems, the BER performance can be improved compared to the conventional schemes.

\section*{Acknowledgment}

This work was supported by the National Research Foundation of Korea (NRF) grant funded by the Korea government (MEST) (No. 2011-0000328).


\begin{thebibliography}{1}




\bibitem{Dae-Woon}
D.-W. Lim, S.-J. Heo, and J.-S. No, ``An overview of peak-to-average power ratio reduction schemes for OFDM signals,'' \textit{J. Commun. Netw.,} vol. 11, no. 3, pp. 229-239, Jun. 2009.

\bibitem{Hyun-Bae}
H.-B. Jeon, K.-H. Kim, J.-S. No, and D.-J. Shin, ``Bit-based SLM schemes for PAPR reduction in QAM modulated OFDM signals,'' \textit{IEEE Trans. on Broadcast.}, vol. 55, no. 3, pp. 679-685, Sep. 2009.

\bibitem{Bauml}
S. H. Muller, R. W. Bauml, R. F. H. Fischer, and J. B. Huber, ``OFDM with reduced peak-to-average power ratio by multiple signal representation,'' \textit{In Annals of Telecommun.}, vol. 52, no. 1-2, pp. 58-67, Feb. 1997.

\bibitem{Seok-Joong}
S.-J. Heo, H.-S. Noh, J.-S. No, and D.-J. Shin, ``Modified SLM scheme with low complexity for PAPR reduction of OFDM systems,'' \textit{IEEE Trans. Broadcast.}, vol. 53, no. 4, pp. 804-808, Dec. 2007.

\bibitem{Wang}
C. L. Wang and Q. Y. Yuan, ``Low-complexity selected mapping schemes for peak-to-average power ratio reduction in OFDM systems,'' \textit{IEEE Trans. Signal Processing,} vol. 53, no. 12, pp. 4652-4660,
Dec. 2005.

\bibitem{Ghassemi}
A. Ghassemi and T. A. Gulliver, ``A low-complexity PTS-based radix FFT method for PAPR reduction in OFDM systems,'' \textit{IEEE Trans. Signal Processing,} vol. 56, no. 3, pp. 1161-1166, Mar. 2008.

\bibitem{Xiaodong}
X. Li and L. J. Cimini, Jr., ``Effects of clipping and filtering on the performance
of OFDM,'' \textit{IEEE Commun. Lett.}, vol. 2, no. 5, pp. 131-133, May 1998.

\bibitem{Ochiai}
H. Ochiai and H. Imai, ``Performance analysis of deliberately clipped OFDM signals,'' \textit{IEEE Trans. Commun.}, vol. 50, pp. 89-101, Jan. 2002.

\bibitem{Donoho}
D. L. Donoho, ``Compressed sensing,'' \textit{IEEE Trans. Inf. Theory}, vol. 52, no. 4, pp. 1289-1306, Apr. 2006.


\bibitem{Candes}
E. Candes, J. Romberg, and T. Tao, ``Robust uncertainty principles: Exact signal reconstruction from highly incomplete frequency information,'' \textit{IEEE Trans. Inf. Theory}, no. 2, pp. 489-509, Feb. 2006.

\bibitem{Candes2}
E. Candes and T. Tao, ``Near-optimal signal recovery from random projections: universal encoding strategies?,'' \textit{IEEE Trans. Inf. Theory}, vol. 52, no. 12, pp. 5406-5425, Dec. 2006.

\bibitem{Wright}
S. J. Wright, R. D. Nowak, and M. A. T. Figueiredo, ``Sparse reconstruction by separable approximation,'' \textit{IEEE Trans. on Signal Processing,} vol. 57, no. 7, pp. 2479-2493, Jul. 2009.

\bibitem{Al-Safadi}
E. B. Al-Safadi and T. Y. Al-Naffouri, ``On reducing the complexity of tone reservation based PAPR reduction schemes by compressive sensing,'' in \textit{Proc. IEEE Globecom 2009}, Honolulu HI, Nov. 2009.

\bibitem{Mohammad}
M. Mohammadnia-Avval, A. Ghassemi, and L. Lampe, ``Compressive sensing recovery of nonlinearity distorted OFDM signals,'' in \textit{Proc. IEEE Int. Conf. Commun.}, Jun. 2011.

\bibitem{Hangjun}
H. Chen and A. Haimovich, ``Iterative estimation and cancellation of clipping noise for OFDM signals,'' \textit{IEEE Commun. Lett.}, vol. 7, pp. 305-307, Jul. 2003.

\bibitem{Dukhyun}
D. Kim and G. L. Stuber, ``Clipping noise mitigation for OFDM by decision-aided reconstruction,'' \textit{IEEE Commun. Lett.}, vol. 3, pp. 4-6, Jan. 1999.

\bibitem{Ying}
Y. Chen, J. Zhang, and A. D. S. Jayalath, ``Estimation and compensation of clipping noise in OFDMA systems,'' \textit{IEEE Trans. Wireless Commun.,} vol. 9, no. 2, pp. 523-527, Feb. 2010.

\bibitem{Tropp}
J. A. Tropp and A. C. Gilbert, ``Signal recovery from random measurements via orthogonal matching pursuit,'' \textit{IEEE Trans. Inf. Theory}, vol. 53, no. 12, pp. 4655-4666, Dec. 2007.

\bibitem{diffset}
D. Gordon, La Jolla Difference Set Repository [Online]. Available: http://www.ccrwest.org/diffsets/diff sets/baumert.html.

\bibitem{Proakis}
J. G. Proakis and M. Salehi, \textit{Communication Systems Engineering}, 2nd ed. Upper Saddle River, NJ: Prentice-Hall, 2002.


\end{thebibliography}
\end{document}